\pdfoutput=1

\documentclass[a4paper,fleqn]{cas-dc}

\usepackage[authoryear,longnamesfirst]{natbib}

\usepackage{fvextra}

\makeatletter
\newenvironment{appendixfigure}
  {\par\medskip\noindent\begin{minipage}{\textwidth}\def\@captype{figure}}
  {\end{minipage}\par\bigskip}
\makeatother

\DefineVerbatimEnvironment{wrapverb}{Verbatim}{
  breaklines=true,
  breakanywhere=true
}

\def\tsc#1{\csdef{#1}{\textsc{\lowercase{#1}}\xspace}}
\tsc{WGM}
\tsc{QE}

\makeatletter
\AtBeginDocument{%
  \let\orig@makefntext\@makefntext
  \long\def\@makefntext#1{\pdfrunninglinkoff\orig@makefntext{#1}\par\pdfrunninglinkon}%
  \g@addto@macro\@floatboxreset{\pdfrunninglinkoff}%
  \@for\@tempa:=figure,figure*,table,table*\do{%
    \expandafter\AtEndEnvironment\expandafter{\@tempa}{\par\pdfrunninglinkon}}%
  \patchcmd{\@outputpage}{\@thehead}{\pdfrunninglinkoff\@thehead\pdfrunninglinkon}{}{}%
  \patchcmd{\@outputpage}{\@thefoot}{\pdfrunninglinkoff\@thefoot\pdfrunninglinkon}{}{}%
}
\makeatother

\begin{document}
\let\WriteBookmarks\relax
\def\floatpagepagefraction{1}
\def\textpagefraction{.001}

\shorttitle{}    

\shortauthors{}  

\title [mode = title]{Introducing ReTiNA-Cinema: \textbf{Re}al-\textbf{Ti}me \textbf{N}euro-\textbf{A}ugmented Cinema via Generative AI}  



%

\author[1,2]{Antoine Bellemare-Pepin}[orcid=0000-0001-7340-1700]

\cormark[1]

\fnmark[1]

\ead{antoine.bellemare9@gmail.com}


\credit{}

\affiliation[1]{organization={CoCo Lab, Department of Psychology, Université de Montréal},
           city={Montreal},
           state={QC},
           country={Canada}}

\author[1,2]{Philipp Thölke}[orcid=0000-0002-3208-2501]

\fnmark[1]

\ead{phil.thoelke@gmail.com}


\credit{}

\affiliation[2]{organization={Kairos Hive},
           city={Montreal},
           state={QC},
           country={Canada}}

\author[1,2]{Yann Harel}[orcid=0000-0002-8970-1983]

\ead{yharel109@gmail.com}

\author[1,2,3]{Mar Estarellas}[orcid=0000-0001-9566-3066]

\ead{marestarellas@gmail.com}

\author[1,3,4]{Karim Jerbi}[orcid=0000-0002-3790-9651]

\ead{karim.jerbi@umontreal.ca}

\affiliation[3]{organization={Department of Psychiatry, McGill University},
           city={Montreal},
           state={QC},
           country={Canada}}

\affiliation[4]{organization={Mila (Quebec AI research Institute), Université de Montréal},
           city={Montreal},
           state={QC},
           country={Canada}}

\affiliation[5]{organization={UNIQUE Center (Quebec Neuro-AI research Center), Université de Montréal},
           city={Montreal},
           state={QC},
           country={Canada}}

\cortext[1]{Corresponding author}

\fntext[1]{These authors contributed equally to this work.}

\ExplSyntaxOn
\cs_set:Npn \__first_footerline: { \mbox{} }
\ExplSyntaxOff


\begin{abstract}
Generative artificial intelligence (AI) is currently revolutionizing content creation across various media. Between those who fear it will replace human creativity and those who regard it solely as a tool, many in the creative industries are beginning to view generative AI as a collaborator. Yet interaction with these systems often relies on discrete, turn-based prompting that can interrupt flow and constrain expression to symbolic text. We explore physiological signals as a source of continuous adaptation without explicit commands, as a path toward embodied co-creation. This paper presents ReTiNA-Cinema, a methodological proof-of-concept that integrates electroencephalography (EEG) with generative AI to enable physiologically guided visual modulation of cinematic material. By linking physiological activity to the evolving appearance of a film, the system provides a basis for exploring embodied sense-making, in which viewers' bodily responses contribute to shaping the sensory experience through which they interpret the unfolding scenes. The system combines real-time EEG signal processing with a generative AI pipeline based on Stable Diffusion Turbo for frame-level image transformation. Single-user and two-user configurations map EEG-derived features to prompt selection and transformation strength. ReTiNA-Cinema offers a framework for exploring how the physiology of viewers might influence a film's visual presentation, extending the possibilities for participation in cinematic experiences and storytelling. This system raises questions about the effects on engagement, perceived agency, and emotional response, which remain open to future empirical study. Video illustrations: \url{https://www.youtube.com/playlist?list=PLMu36WzSQKiVeBnrUdwUAoUhqLqGX3_bw} and implementation: \url{https://github.com/KairosHive/retina-cinema}.
\end{abstract}




\maketitle


\section{Introduction}\label{sec1}

Recent advances in generative artificial intelligence (AI) have significantly expanded the capabilities of content creation across various modalities including textual~\citep{vaswani_attention_2023}, visual~\citep{ho_denoising_2020} and auditory media~\citep{kreuk_audiogen_2023}. Most notably, the recent rise of diffusion models for image generation has demonstrated remarkable proficiency in producing high-quality, realistic content ~\citep{rombach_high-resolution_2022, esser_scaling_2024}. These technological breakthroughs have propelled generative AI to become an important tool for creative expression in digital arts, partly because they enable novel forms of real-time interactivity with high-grade visual material.\\

Despite these advancements, most creative interaction with generative AI remains discrete and turn-based: users issue a prompt, the system responds, and the user iterates with a new prompt~\citep{dang_how_2022}. This stop–start rhythm imposes attentional switching costs, increases working-memory load, and fragments creative flow, particularly in time-sensitive media like film where the experience unfolds continuously. It also bottlenecks expression into typed text, which can be ill-suited for tacit, affective, or pre-verbal states that shape aesthetic intent.\\

To address this limitation, we propose to take advantage of an electroencephalography (EEG) bio-adaptive interface to interact fluidly with generative AI. While bio-adaptive art is an emerging field, it is made increasingly accessible through the use of mobile electrophysiology devices and the improved control over the output of generative AI models. Previous systems have explored EEG-driven audiovisual generation and biofeedback-based visual experiences~\citep{tholke_bio-mechanical_2024, kitson_lucid_2019}, motivating further investigation of physiological input in creative interaction. We present this work as a methodological proof-of-concept whose objective is to map EEG-derived features to prompt selection and transformation strength during generative processing of existing film frames. Accordingly, our design choices were guided by three practical constraints that also define the main strengths of the system: (i) supporting continuous EEG-driven visual updates with low processing latency, (ii) maintaining an interpretable mapping between EEG features and visual transformations, and (iii) using a modular architecture that can be extended in future empirical work. While such systems have the potential to promote engagement and enable richer interaction, we do not evaluate those outcomes in this study. Instead, this work establishes a foundation for future empirical studies.\\

\subsection{Why bio-adaptive? User needs and theory.} 
Three potential benefits motivate the exploration of bio-adaptive interfaces in creative media:
(i) Lower interaction overhead \& support flow. Research on flow and creative performance motivates reducing interaction demands during creative activities~\citep{dietrich_neurocognitive_2004, pacauskas_information_2017}. Continuous physiological control may reduce the need for repeated manual prompting, although its effects on attention, engagement, and perceived agency in this application remain to be evaluated.
(ii) Express tacit/affective intent. Physiological signals correlate with arousal, engagement, and cognitive dynamics; mapping these to cinematic transformations provides a continuously varying input for visual adaptation without requiring explicit commands. \citep{shu_review_2018, ji_characterizing_2024}.
(iii) Enable accessibility and co-creation at scale. Bio-adaptive control can complement or substitute explicit input (hands/voice) in performance and shared viewing scenarios, with potential extensions to multi-user coupling and audience-level modulation.
Grounded in enactive and participatory sense-making~\citep{de_jaegher_participatory_2007, van_dijk_designing_2016}, ReTiNA-Cinema treats the viewer’s cognitive dynamics as part of the media’s control loop, aligning the temporal grain of internal state with the temporal grain of the film, which discrete prompting cannot easily match.

\subsection{Cinema as Context for Bio-Adaptive Interaction}
Specifically, in this paper, we present an application of the broader generative AI framework introduced above, focusing on its implementation in the field of cinema. Cinema serves as a particularly compelling context for exploring the integration of bio-adaptive art with existing media. Historically, cinema has been situated at the forefront of new technologies, pushing the state of the art in areas such as computer-generated imagery (CGI) and 3D animation to enhance storytelling and audience engagement~\citep{mendiburu_3d_2009, watkins_3d_2000}. Generative AI's ability to transform existing video content introduces new opportunities for adaptive storytelling in cinema. Movies inherently engage viewers at an intuitive layer, making them an ideal medium for the exploration of embodied human-AI co-creation. By incorporating brain signals to modulate cinematic visuals dynamically, we aim to connect the viewer's physiological responses with the film's evolving appearance, creating a bio-adaptive experience with the potential to support immersion. We call such systems Real-Time Neuro-Augmented (ReTiNA), a concept that can in principle apply to various media forms including but not limited to cinema.\\

ReTiNA aims at transforming movie-watching into an adaptive experience, thus encouraging a creative synergy between human and AI. This allows users to actively shape the visual experience, while also passively perceiving narratives that are dynamically adjusted. We further demonstrate a two-user configuration that combines features from two EEG streams to adapt the visual presentation of the film. ReTiNA represents an early exploration of embodied human-AI co-creation in cinema. In the single-user configuration, EEG-derived features select between two prompts and modulate the strength of real-time transformations applied to incoming movie frames, producing an evolving visual interpretation of the film. ReTiNA-Cinema frames generative AI as an \emph{active collaborator} embedded within a closed loop. Rather than replacing human creativity, the system continuously interprets and responds to users’ cognitive dynamics, distributing agency across human and machine in ways that neither could achieve alone. This collaborative stance points toward a new mode of human–AI relations, where generative systems act as co-participants that extend human sense-making in real time rather than standing apart as tools to be used or competitors to be feared. This work lays the groundwork for future explorations into decoding latent representations from the brain, potentially eliminating the need for predefined textual prompts and paving the way for multimodal bio-adaptive media experiences.

\section{Related Works}
Generative AI has been rapidly adopted across creative practice and film for ideation, script drafting/revision, style transfer, and video/image augmentation, accelerating iteration throughout the pipeline \citep{rombach_high-resolution_2022, esser_scaling_2024}. However, most deployments remain \emph{prompt-driven and turn-based}, which optimizes throughput yet interrupts creative flow in time-based media where the experience unfolds continuously \citep{roush_llm_2023, huang_filmaster_2025}. Within interactive and adaptive cinema, audience-in-the-loop works have tended to rely on explicit branching or discrete control, producing pauses or narrative stalls that can break immersion (e.g., \emph{Late Shift}, 2016; \emph{Black Mirror: Bandersnatch}, 2018). Early systems showed that adding user control can increase engagement---e.g., \emph{DeepSea Adventure} \citep{hu_user_2005}---but interaction was still turn-based. Affective and physiological approaches often focused on \emph{classifying} emotional states offline or at coarse timescales without continuously modulating visuals in real time \citep{oliveira_towards_2011}, or they routed physiology to \emph{non-visual} channels such as adaptive audio or narrative branching without frame-level image transformation (e.g., VFC \citep{bellemarevfc_2024}). In parallel, bio-adaptive artworks and installations have mapped EEG and other biosignals to audiovisual transformations to evoke embodied engagement \citep{kitson_lucid_2019, tholke_bio-mechanical_2024}, but typically outside the context of feature-length cinema and without state-of-the-art frame-rate generative pipelines.
Our contribution builds a bridge between these lines of work: we couple an EEG feature extractor to modern diffusion models and operate them in a \emph{continuous, closed-loop} configuration that performs real-time image-to-image transformations of \emph{existing} cinematic material at sub-second latency. This enables high-rate, viewer-state–driven modulation of the same cinematic substrate---without branching or narrative pauses---supporting both single- and two-user settings. In doing so, we connect affective/bio-adaptive interaction with contemporary generative pipelines specifically for \emph{cinema}, rather than standalone visualizations, and we position the system as a proof-of-principle platform for future empirical studies comparing and combining EEG with other modalities.

\section{System Design}\label{sec2}

\begin{figure*}
    \centering
    \includegraphics[width=\textwidth]{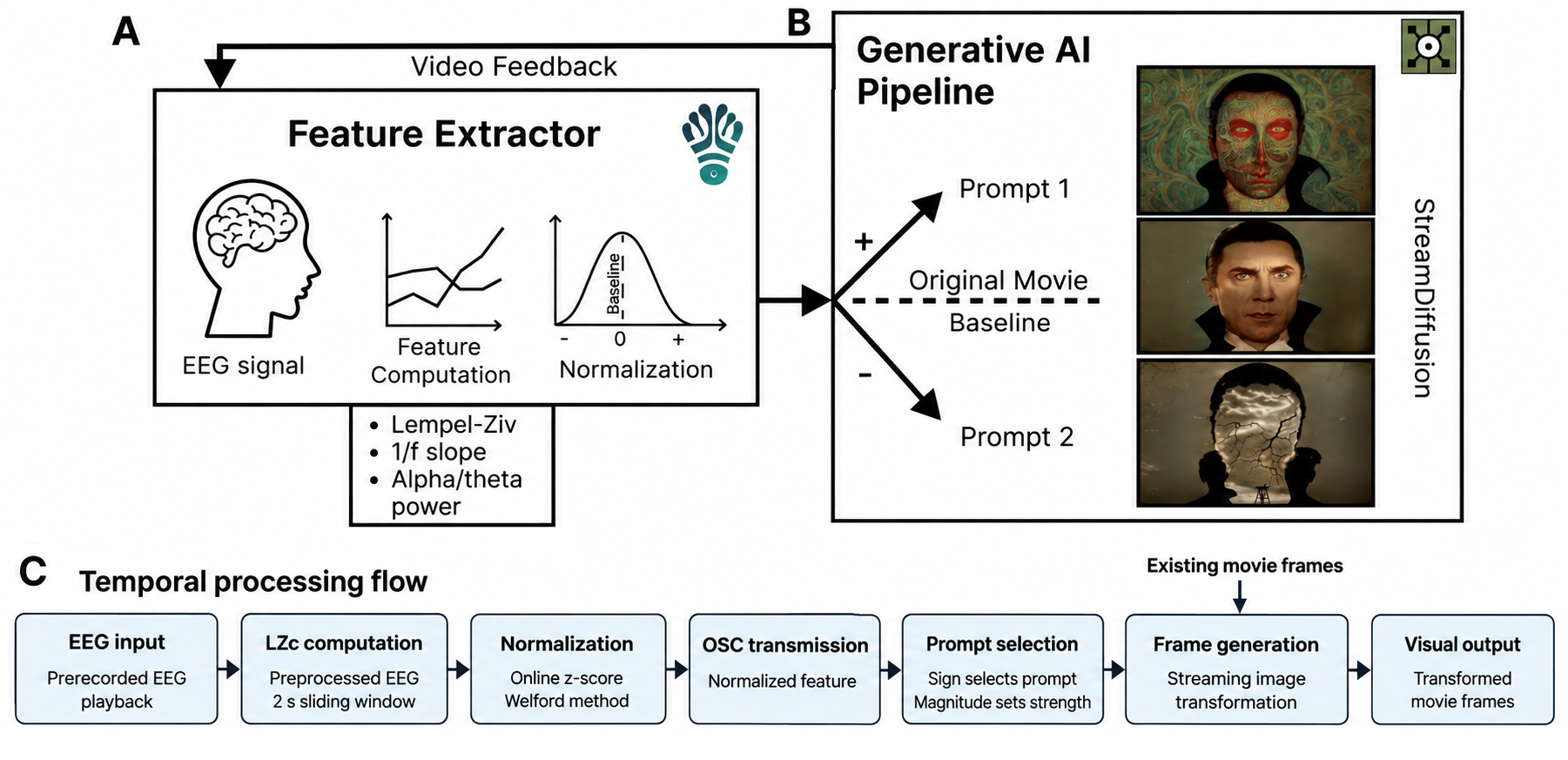}
    \caption{Architecture of our ReTiNA-Cinema pipeline showing the EEG feature extractor and generative AI modules, highlighting the continuous user feedback loop. In panel A, band-power and 1/f-slope features indicate supported alternatives that were not used in the reported demonstrations; LZc was the feature used for single-user visual control. Source: Authors.}
    \label{fig:neuro-augmented-cinema-figure}
\end{figure*}

The proposed system integrates a bio-adaptive interface with real-time generative AI to create an adaptive augmented cinema experience. The current implementation modifies the visual appearance of film frames without explicitly editing plot, dialogue, or scene order. These transformations may nevertheless influence how viewers perceive and interpret the narrative. It comprises two core components: the EEG feature extractor and the generative AI pipeline, designed for real-time experience with a sub-second latency (see Figure~\ref{fig:neuro-augmented-cinema-figure}). This ensures a seamless interface between the user and the content of the adaptive movie. It supports both single-user and two-user configurations, enabling personalized and collaborative interactions. The repository required to run the full system is publicly shared to facilitate reproducibility and further research; see \href{https://github.com/KairosHive/retina-cinema}{github.com/KairosHive/retina-cinema}.\\

In this paper, a single EEG feature, Lempel-Ziv complexity (LZc), is employed to modulate the system's feedback. LZc has been widely used in research to quantify signal complexity and capture various cognitive states~\citep{mateos_measures_2018, aboy_interpretation_2006, zhang_normalized_2009, farnes_increased_2020, orlowski_sensory_2023, lewis-healey_breathwork-induced_2024, starkey_statistical_2023}. While we recognize the inherent challenges in defining mental states based on a single feature, this decision was made to prioritize clarity and keep the feature-to-visual mapping simple and explicit. It is important to note, however, that other features could be integrated into the system in future work to capture a different or broader range of cognitive dynamics.\\

\begin{figure}
    \centering
    \includegraphics[width=0.5\textwidth]{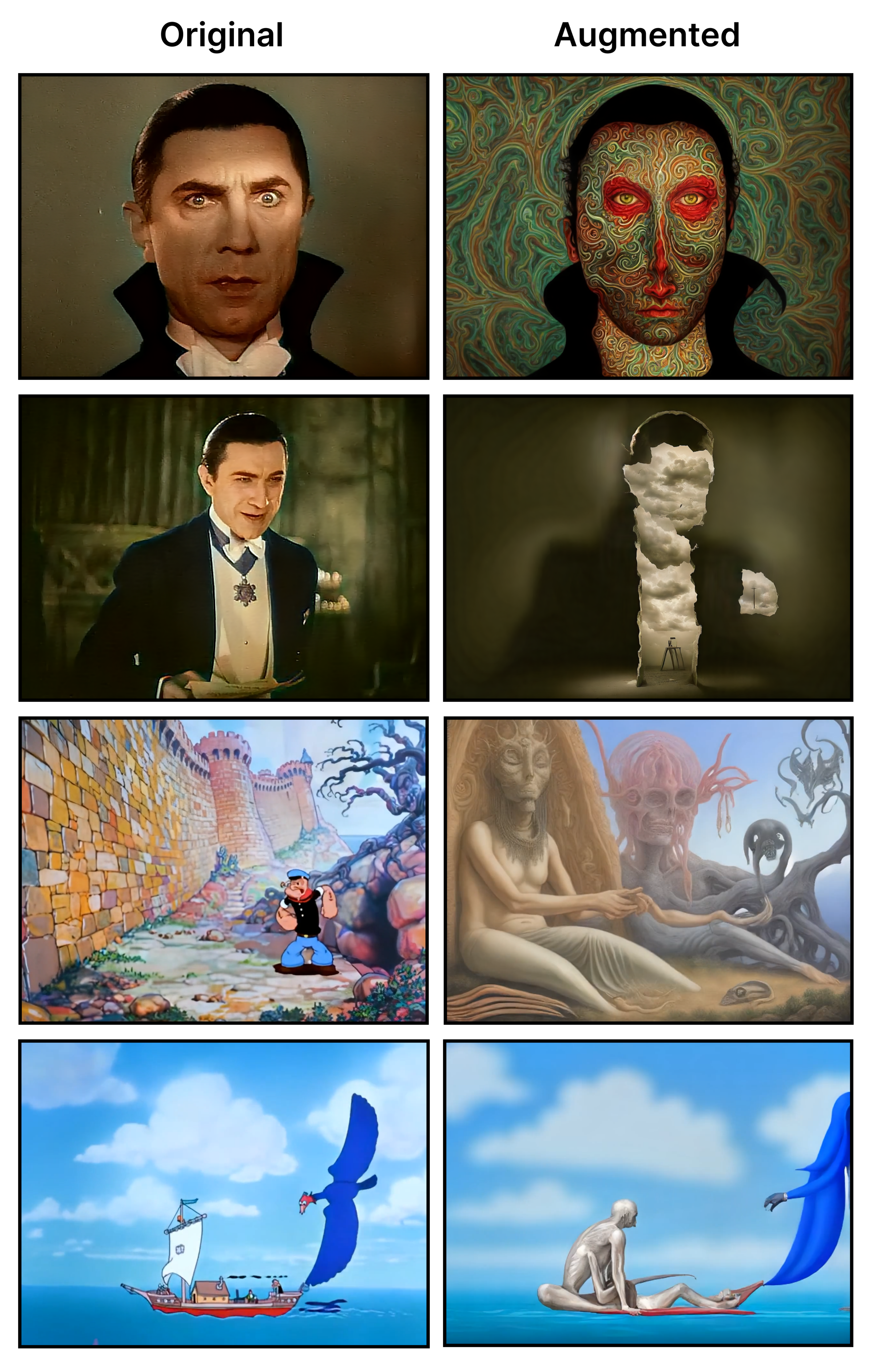}
    \caption{Examples of visual transformations produced by ReTiNA-Cinema. Left: original frames from Frankenstein (1931; top two rows) and Popeye the Sailor Meets Sindbad the Sailor (1936; bottom two rows). Right: corresponding AI-transformed frames produced by the authors. Original film frames: [insert verified copyright attribution and reproduction basis for each source]. Transformations and figure assembly: Authors. Source: Authors.}
    \label{fig:RETINA_examples}
\end{figure}

Stills captured from a recorded video demonstrating the current system’s output are shown in Figure~\ref{fig:RETINA_examples}. These visual examples showcase the variety of transformations achieved through EEG-driven interactions, illustrating the potential of the system to create dynamic, personalized cinematic experiences.

\subsection{EEG Feature Extractor}

The EEG feature extractor (see Figure~\ref{fig:neuro-augmented-cinema-figure}A) is implemented in \textit{goofi}, an open-source framework designed for real-time biosignal processing (\href{https://github.com/KairosHive/goofi}{github.com/KairosHive/goofi}). This platform offers both flexibility and extensibility, making it suitable for a wide range of research-creation applications. The system supports multiple EEG devices through the Lab Streaming Layer (LSL) protocol~\citep{kothe_lab_2024}, accommodating a wide range of devices from consumer-grade headsets (e.g. Muse, OpenBCI) to high-end research equipment (e.g. DSI-24, BitBrain). A screenshot of our implementation in \textit{goofi} can be found in Supplementary Figure 1.

\paragraph{Why EEG?}
EEG provides a hands-free, moment-to-moment signal that can reflect covert attention and engagement while viewing—without asking the viewer to act. Compared to ECG (slower, autonomic arousal; less specific to cognitive state) and facial expression (overt, culturally biased, and often decoupled from felt experience in cinema), EEG tracks the fast, internal dynamics that shape viewing—fluctuations in attention, engagement, and cognitive effort—at the timescale of film. Recent EEG foundation models also improve robustness and enable lightweight inference of proxies for attention/engagement from short windows. Our use here is proof-of-principle; the pipeline is modality-agnostic and suited to future studies that compare EEG with ECG/facial cues or fuse them for stronger bio-adaptive control.

\subsubsection{Signal Pre-processing}

To ensure the integrity and reliability of EEG data, signal pre-processing is a critical step. We apply a causal fourth-order Butterworth band-pass filter ranging from 1 to 40 Hz to remove slow drift and high-frequency noise artifacts, such as muscle movements and electrical interference, while retaining genuine brain oscillations~\citep{khosla_comparative_2020}. The filter state is carried across sample blocks so that each sample is filtered exactly once, and the first two seconds of output are discarded to suppress the filter's start-up transient. The filtered stream is then accumulated in a two-second rolling buffer that discards any signal older than two seconds. The system ingests EEG at any sampling rate; our demonstration used 64-channel EEG sampled at 1024 Hz, so the buffer holds 2048 samples per channel.

\subsubsection{Feature Extraction}\label{sec:feature-extraction}

In this work, the primary metric for modulating cognitive states is Lempel-Ziv Complexity (LZc), computed over the two-second rolling buffer to balance robustness with responsiveness. We use a normalized variant of LZ76~\citep{zhang_normalized_2009} with mean binarization, calculated per EEG channel, and then averaged across the scalp. LZc estimates are biased and highly variable for short sequences, so the window must be long enough to give a stable count. At the same time, the window must be short enough that a change in brain state is reflected on screen within a few seconds, which is the time scale of a cinematic cut or gesture. The feature is recomputed every time a new block of EEG arrives, 30 times per second. Since the window advances by only about 34 samples per update, consecutive windows share 2014 of their 2048 samples, an overlap of 98.3\%, and a window is fully replaced only after 60 updates. The 30 Hz feature stream should therefore be read as a smoothly interpolated trajectory of the underlying two-second estimate rather than as 30 independent measurements per second: successive values are strongly autocorrelated. To automatically adjust the dynamic range of the feature, we z-transform the complexity feature using Welford's method~\citep{welford_note_1962}. This adaptive normalization controls for inter-individual variability in brain states and dynamically adjusts to fluctuations in recording conditions, ensuring stable performance and a personalized experience across users.\\

LZc is a well-established information-theoretic measure quantifying the diversity and complexity of EEG signals. Empirical research has linked higher LZc values to states of alert wakefulness, increased cognitive flexibility, and altered states such as those induced by breathwork or psychedelics ~\citep{li_cortical_2019, hohn_spectral_2024, timmermann_neural_2019, lewis-healey_breathwork-induced_2024, obyrne_how_2022, tholke_caffeine_2025}. These associations provide a foundation for using LZc as a proxy for the viewer’s current neural dynamism—capturing moments when cognitive processes may be more exploratory or enriched, which in turn drives the dynamic visual transformations. Although no single EEG feature can fully encapsulate the multifaceted nature of mental states, LZc offers valuable insights into the overall complexity of brain activity, differentiating it from random or noisy data by reliably correlating with known states of consciousness.\\

In the two-user configuration, the system is designed to enable the selection of various metrics to assess inter-brain coupling, with the weighted phase-lag index (wPLI;~\citep{vinck2011improved}) being used in the current work. We utilize the same filtered rolling buffer that was used for LZc calculation. We restrict the channel selection to occipital sensors only to focus inter-brain synchrony on shared visual experience. The inter-brain coupling is then derived as the average wPLI in the alpha band (8-12 Hz) via the mean over pairs with one channel from each participant, excluding within-participant pairs. The resulting value is z-scored against running (Welford) statistics accumulated over the session.\\

The weighted phase-lag index (wPLI) quantifies phase synchronization between signals~\citep{vinck2011improved} and has been used to investigate inter-brain synchronization during turn-taking verbal interaction~\citep{ahn_interbrain_2018}. These findings motivate its use in exploring interpersonal interaction through adaptive media, although wPLI alone does not establish shared cognitive states. When two viewers receive the same audiovisual input, inter-brain wPLI may partly reflect responses to the shared stimulus rather than interpersonal interaction or shared cognition. We also support a range of other measures of inter-brain synchrony, such as coherence~\citep{balconi_dyadic_2023}, to provide alternative coupling measures for the two-user configuration.\\

Importantly, our system is designed as a proof-of-concept. The modular feature extraction architecture also supports spectral power in canonical frequency bands and the slope of the aperiodic (1/f-like) spectral component. These features were not used in the reported demonstrations. LZc was used for single-user visual control; the two-user configuration additionally used wPLI to control transformation intensity. By demonstrating how LZc can drive visual transformations in real time, this work establishes a modular platform for exploring alternative or complementary EEG features. Future studies could investigate whether these features support more specific mappings to cognitive or emotional states, subject to independent validation.

\subsection{Generative AI Pipeline}

The generative AI pipeline dynamically transforms movie visuals in real time, guided by the EEG-derived features (see Figure~\ref{fig:neuro-augmented-cinema-figure}B). Using predetermined prompts and state-of-the-art AI models, the system continuously modulates the cinematic experience to generate an adaptive and responsive visual narrative.

\subsubsection{Platform and AI Engine}

The pipeline is implemented using \textit{TouchDesigner}, a node-based visual programming environment ideal for real-time multimedia content~\citep{derivative_touchdesigner_nodate}. The AI engine utilizes the Stable Diffusion Turbo model~\citep{sauer_adversarial_2023} within the StreamDiffusion framework~\citep{kodaira_streamdiffusion_2023}. We use StreamDiffusion as the real-time backbone for image-to-image because it treats generation as a \emph{stream} rather than a sequence of isolated images. Concretely, it (i) keeps models and intermediate buffers resident on the GPU to avoid per-frame reinitialization, (ii) exposes low-latency controls (denoise strength, guidance, prompt weights) that we can update every frame from EEG, and (iii) reuses state across frames to reduce overhead and improve temporal stability at the same compute budget. This configuration enables fast, diffusion-based image-to-image transformation of movie frames, achieving approximately 15 frames per second on hardware such as the RTX 4090 (Laptop) and RTX 3090 (Desktop). The architecture is model-agnostic and upgradeable: higher-quality diffusion checkpoints can be swapped in as hardware permits, though some newer architectures are not yet stable for streaming image-to-image; accordingly, model choice reflects the speed–quality trade-off required to preserve real-time frame rates. A screenshot of our implementation in TouchDesigner can be found in Supplementary Figure 2.

\subsubsection{Prompt Engineering and Selection}

The system provides two complementary modes of prompt preparation: (i) manual user-defined prompts and (ii) dynamically generated, context-aware prompts created by a large language model (LLM). \\

In the manual mode, users define two textual prompts for the single-user configuration or four for the two-user configuration before movie playback. These prompts represent different aesthetic or thematic transformations that alternate dynamically based on the user’s brain state. For example, in the single-user configuration, a viewer may choose Prompt~1 as \texttt{"abstract painting, pastel textures"} and Prompt~2 as \texttt{"glitch art, fragmented structures, high contrast neon"}. This dual-prompt structure enables expressive transitions between visually distinct interpretations of the same cinematic content, supporting contrastive modulation of visuals in real time. Users may enter custom prompts directly in TouchDesigner to tailor the experience to their chosen film.\\

In the dynamic mode, prompts are generated automatically during the movie based on periodic visual context. At fixed intervals (default: every 30 seconds, user-adjustable), the system captures a still frame from the movie and submits it to a visual LLM (either locally hosted or via API). The LLM analyzes the frame and generates four complementary visual style descriptions that are designed to be orthogonal in mood, atmosphere, and artistic perspective. To ensure interpretable neurofeedback, the system constrains the prompt-generation process so that: (i) \emph{Style~1} is always constructive (adding styles, details, and complexity), (ii) \emph{Style~4} is always deconstructive (reducing complexity, minimalism, abstract), while (iii) \emph{Styles~2 and 3} remain unconstrained. In the single-user configuration, Style~1 and Style~4 serve as Prompt~1 and Prompt~2, respectively. The two-user configuration uses all four styles. This mapping guarantees a clear alignment between EEG-derived features and visual transformations, providing an intuitive and consistent bio-adaptive control signal. \\

The exact instruction prompt used for LLM-based prompt generation is provided in the Supplementary Materials, ensuring transparency and reproducibility of this method.

\subsubsection{Single-User Interaction}

In the single-user configuration, the normalized LZc from the EEG data determines both the selection of the prompt and the strength of the visual transformation of the movie frames. According to the normalization strategy, the feature values are centered around zero (baseline), with values above and below the baseline represented as positive and negative deviations, respectively (see Figure~\ref{fig:neuro-augmented-cinema-figure}B). This means that the amplitude by which the feature deviates from the subject-normalized average is what drives the intensity of the modulation. Specifically:

\begin{itemize}
    \item Positive normalized values activate the first prompt.
    \item A value of zero (baseline) results in displaying the original movie without any effects.
    \item Negative normalized values trigger the second prompt.
\end{itemize}

The intensity of the effect is smoothly modulated in proportion to the magnitude of the normalized feature value. This design aims to provide an intuitive and responsive experience, allowing EEG-derived variations in brain activity to influence the visual appearance of the cinematic content in real time.

\subsubsection{Two-User Interaction}

In the two-user configuration, the system operates in two parallel streams: the first stream determines the intensity of the visual transformation, while the second stream controls the selection of prompts that shape the overall aesthetic or thematic direction of the transformation.\\

The chosen feature for inter-brain synchrony, in this case wPLI~\citep{vinck2011improved}, modulates the intensity of movie transformations. The use of the wPLI score follows the same principle as in the single-user setup, i.e. the normalized average cross-brain coupling is mapped onto the diffusion model's denoising strength parameter, resulting in a continuously varying intensity of the effect. The system is not limited to wPLI; other neurophysiological metrics such as coherence or mutual information could also be used depending on the desired interaction dynamics. A future multi-user extension could aggregate features across more than two EEG streams to control transformation strength, enabling group-level interactions in larger collaborative environments.\\

For prompt selection in two-user interactions, the system chooses from four distinct prompts (either user-defined or dynamically generated via a context-aware visual LLM) based on the combination of the normalized LZc values from both users:

\begin{itemize}
    \item \textit{Prompt~1}: Both users exhibit positive feature values.
    \item \textit{Prompt~2}: User~1 has a positive and User~2 has a negative feature value.
    \item \textit{Prompt~3}: User~1 has a negative and User~2 has a positive feature value.
    \item \textit{Prompt~4}: Both users have negative feature values.
\end{itemize}

This \verb|(++, +-, -+, --)| mapping is intentionally simple and taxonomy-agnostic: prompts are either user-defined or scene-informed by a vision LLM, avoiding assumptions about a specific cognitive model. Table~\ref{tab:single_vs_multi_simple} summarizes the differences between the single-user and two-user configurations.\\

The system is implemented and operational for single- and two-user modes, and conceptually extends to larger groups by aggregating coupling across all pairs, but the current implementation (single inter-brain block, four-way sign rule, two OSC streams) is limited to at most two participants.

\begin{table*}
\caption{Single-user and two-user configurations.}
\label{tab:single_vs_multi_simple}
\begin{tabular*}{\tblwidth}{@{}LLL@{}}
\toprule
\textbf{Aspect} & \textbf{Single-user} & \textbf{Two-user} \\
\midrule
EEG inputs
& One EEG stream; LZc
& Two EEG streams; LZc + wPLI \\

Prompt source
& User-defined \emph{or} scene-informed (vision LLM)
& User-defined \emph{or} scene-informed (vision LLM) \\

Prompt set
& 2 prompts
& 4 prompts \\

Selection rule
& Polarity of LZc: $\{+, 0, -\} \rightarrow \{\mathrm{P1}, \mathrm{orig}, \mathrm{P2}\}$
& User pair polarity pattern: $\{++, +-, -+, --\} \rightarrow \{\mathrm{P1}-\mathrm{P4}\}$ \\

Intensity control
& $|\mathrm{z}(\mathrm{LZc})|$ modulates img2img strength
& wPLI (or group average) modulates strength \\

Performance
& $\sim$15 FPS; sub-second updates
& $\sim$15 FPS; sub-second updates \\
\bottomrule
\end{tabular*}
\end{table*}

\subsection{Real-Time Interaction}

The EEG feature extractor updates at 30 Hz: every 33 ms the two-second rolling window advances by the newest samples and LZc (and, in the two-user configuration, wPLI) is recomputed over it. Each update is sent to the generative AI pipeline over the Open Sound Control (OSC) protocol, where it continuously modulates prompt selection and transformation strength in the incoming video stream. Because OSC is widely supported in digital creative tools, the same feature stream can also drive audio effects or generative music, opening the pipeline to multimodal experiences.\\

\begin{figure*}
    \centering
    \includegraphics[width=\textwidth]{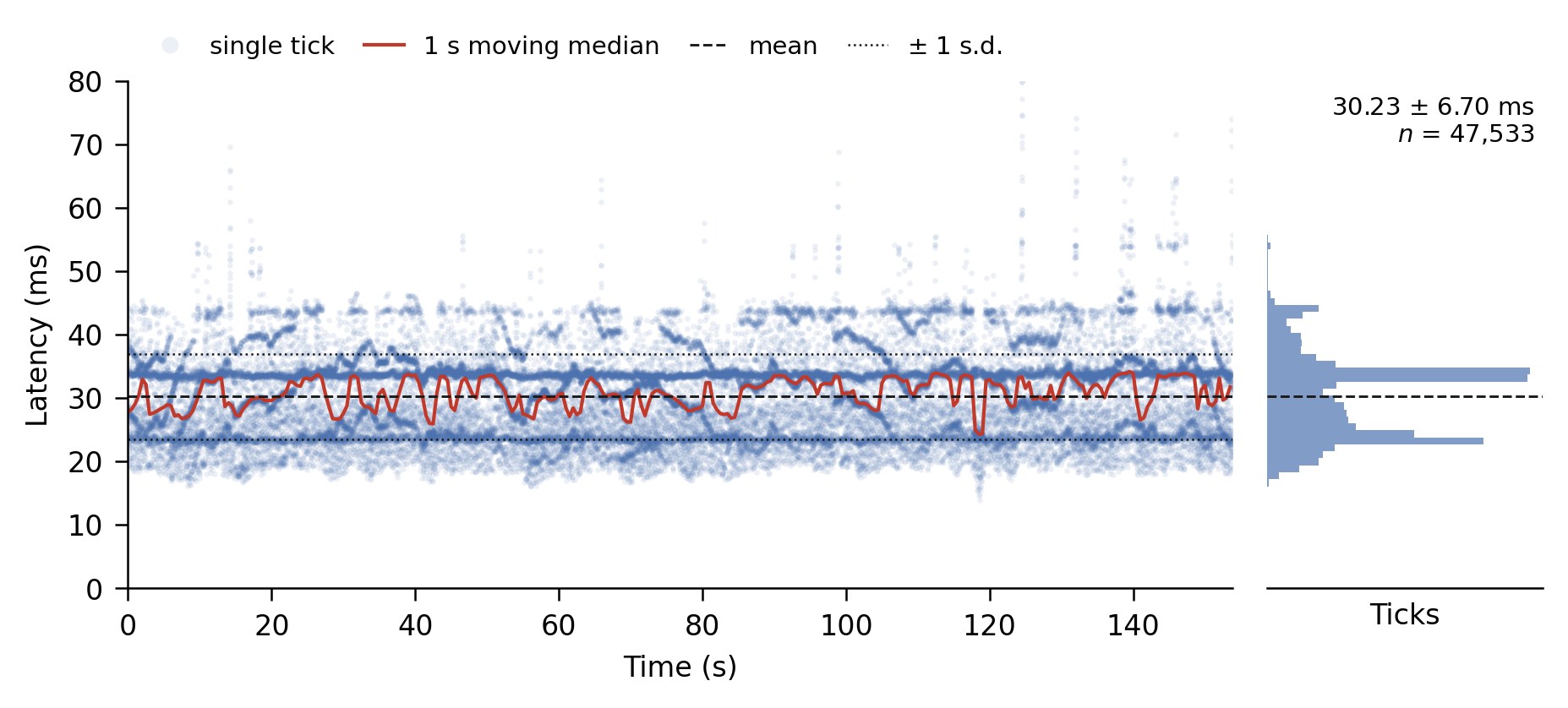}
    \caption{Latency measurement of real-time feature extraction pipeline in goofi including data acquisition, re-referencing, preprocessing, buffering, LempelZiv complexity, cross-subject wPLI and feature normalization. Source: Authors.}
    \label{fig:latency}
\end{figure*}

We measured the end-to-end latency of the feature extractor in \textit{goofi} as the time from the arrival of an EEG sample block at the acquisition node to the emission of the corresponding normalized feature at the OSC output node (Figure~\ref{fig:latency}). Over a 154~s recording, latency was 30.23 $\pm$ 6.70 ms (mean $\pm$ s.d.; range 14--80 ms). Most of this time is spent in the LZc computation (21.6 $\pm$ 3.4 ms), followed by assembling the OSC feature table (7.0 $\pm$ 5.5 ms); acquisition, re-referencing, pre-processing, buffering and normalization together add less than 2 ms on average. The remaining stages of the loop add further latency. OSC transport between \textit{goofi} and \textit{TouchDesigner} on the same machine, which is typically well below one millisecond, and the image-to-image generation in \textit{StreamDiffusion}, which at the observed $\sim$15 frames per second using an NVIDIA RTX 4090 laptop graphics card (Table~\ref{tab:single_vs_multi_simple}) bounds the visual update interval to roughly 70 ms. The total time from a change in the EEG to a change on screen is therefore on the order of 100--150 ms, well below the sub-second scale at which the two-second LZc window itself evolves. While the frame rate of the generative visuals appeared stable during operation, frame-time variability and dropped-frame rates were not systematically characterized in the present evaluation and remain to be quantified in future work.\\

This continuous sub-second update, rather than discrete prompt cycles, provides the temporal continuity required for closed-loop interaction. As discussed in Section~\ref{sec:feature-extraction}, consecutive windows overlap by 98.3\%, so the 30 Hz stream is a smooth trajectory of a two-second estimate; the update rate governs how quickly a change begins to show on screen, while the window length governs how quickly it fully develops.

\begin{sloppypar}

\subsection{Reproducibility}
For reproducibility, the implementation reported here used \textit{goofi} (\href{https://github.com/KairosHive/goofi}{github.com/KairosHive/goofi}) at commit \texttt{550e6c81631d0a742cbf954a8219fffad94bbcd5}, StreamDiffusion version 0.2.6 (\href{https://www.patreon.com/dotsimulate/posts/streamdiffusiont-122151912}{patreon.com/dotsimulate/posts/streamdiffusiont-122151912}), and TouchDesigner version 2025.31550. These versions correspond to the software configuration used for the reported system demonstrations and performance measurements.
\end{sloppypar}



\section{Discussion}\label{sec12}

This paper introduces a system that uses EEG signals to modulate cinematic visuals through generative AI, offering a bio-adaptive viewing experience. The system enables users to influence the movie in real time based on their cognitive states, and also supports collaborative interactions through inter-brain synchrony in the two-user configuration. This work contributes to the emerging field of adaptive cinema by incorporating real-time brain-driven interactions that allow viewers to actively shape their cinematic experience. A playlist with video recordings of the system in action can be found at \href{https://www.youtube.com/playlist?list=PLMu36WzSQKiVeBnrUdwUAoUhqLqGX3_bw}{this YouTube playlist}. 

\subsection{Exploring New Forms of Agency}

The integration of bio-adaptive interfaces with generative AI offers a basis for exploring user agency, potentially allowing individuals to shape their movie experience through dynamic variations informed by EEG correlates of cognitive states. This approach reduces the need for repeated manual prompting and may limit attentional switching, while allowing changes in EEG associated with affective dynamics to indirectly shape the unfolding visual experience.\\

A live implementation could support both implicit and explicit forms of user control. Implicit control emerges as the user's brain activity naturally modulates the generative AI output without deliberate effort, providing a possible basis for implicit interaction. Explicit control builds on neurofeedback research showing that individuals can learn to intentionally modulate aspects of their EEG activity, such as alpha oscillations~\citep{ros_endogenous_2010}. This provides a basis for exploring deliberate control of visual transformations. These potential modes of interaction could accommodate different preferences and levels of involvement. Changes in EEG-derived features shape the visual experience without requiring users to consciously interpret or control those signals. Depending on the selected prompts, visual transformations can range from surreal or abstract imagery to calmer, more familiar aesthetics, creating different visual interpretations of the unfolding film. If users learn to influence the controlling features, the same mapping could also provide a means of deliberately steering these transformations.\\

Both modes place the EEG signal within a broader creative process: the prompts define the aesthetic possibilities, the EEG-derived features select and modulate them, and the generative model produces the resulting imagery. This division of roles provides a basis for the ``AI-as-collaborator'' perspective from computational creativity and HCI. Control is \emph{distributed} across the human choices that configure the experience, the EEG-driven modulation, and the model's generative process. The resulting visual interpretation is therefore neither specified entirely by the user nor generated independently of human input.\\

In addition to user-defined prompts, the system's dynamic prompting mode introduces another potential dimension of agency: by generating context-aware prompts from movie frames, the LLM continuously refreshes the available aesthetic trajectories. This extends implicit control beyond fixed contrasts, allowing neural signals to navigate an evolving set of stylistic possibilities rather than oscillating between two pre-selected transformations. This design motivates the exploration of agency as both distributed (between user and AI) and generative, as the system co-creates new stylistic affordances over time.

\subsection{The Emerging Field of Adaptive Cinema}

Our work contributes to the emerging field of adaptive cinema, where films dynamically respond to the viewers. Previous approaches to adaptive storytelling have often focused on interactive narratives driven by explicit user inputs and branching decision points~\citep{late_shift_2016, bandersnatch_2018}, limiting the scope of real-time, non-linear engagement. By incorporating generative AI into a bio-adaptive interface, we leverage the inherent complexity of brain signals to explore adaptive narratives that evolve fluidly in response to the viewer's changing mental states. Here, narrative adaptation refers to the potential influence of visual transformations on the perceived mood, meaning, or emphasis of a scene, rather than explicit changes to story structure.\\

This bio-adaptive storytelling approach invites viewers to participate in a process of embodied sense-making, where each viewing of a familiar movie becomes a unique experience, shaped by the viewer's ongoing, dynamic engagement with the cinematic narrative. As EEG signals modulate visual experience, the viewer's brain state is no longer a passive backdrop but an active co-creator in this process, continuously and implicitely influencing the evolving narrative.\\

The interaction is therefore not merely one of input-output between user and system, but a co-determined engagement in which sense-making, in the enactivist sense~\citep{de_jaegher_participatory_2007, van_dijk_designing_2016}, unfolds through lived engagement with an adaptive environment that is itself shaped by the viewer's cognitive and affective states. Unlike traditional embodied interactions in the physical world, this loop is mediated by an algorithmic process that lacks its own embodied history, raising questions about the depth and nature of the resulting experience. Yet the viewer's cognitive processes and the AI's algorithmic responses may contribute to a co-constituted experience of the narrative, transforming neural patterns into an active, participatory element of the cinematic medium and creating a layered dynamic in which human agency, lived experience, and algorithmic processes continuously influence each other.

\subsection{Limitations}

While the system demonstrates promising capabilities, several limitations must be acknowledged. The psychological interpretation of the EEG-derived features remains unvalidated in this application. LZc and wPLI are used as computational control features, rather than direct measures of engagement, emotion, agency, or shared cognition. Evaluation with prerecorded EEG playback demonstrates the feature-to-visual mapping but does not establish these psychological outcomes.\\

Dynamic prompt generation also introduces new challenges: the interpretability and stability of LLM-produced prompts may vary across contexts, raising questions of reproducibility and consistency. While constraining constructive versus deconstructive mappings helps preserve a clear neurofeedback signal, future work must evaluate how such variability influences user perception and agency.\\

We also acknowledge that current outputs sometimes skew toward specific aesthetics, reflecting prompt design and diffusion-model biases, which we treat as a limitation to be addressed through prompt calibration and alternative model checkpoints. Further post-processing can be implemented depending on the desired style.\\

This work does not include an empirical user study. The technical demonstrations use prerecorded EEG playback and therefore do not evaluate viewer responses or a live viewer--system feedback loop. Planned controlled user testing is described in Section~\ref{sec:future-work}.\\

Extending the current two-user configuration to larger groups would require additional aggregation rules, potentially based on averaging or clustering normalized feature values. By mapping inter-brain coupling to visual transformations, the system offers a basis for visually adaptive, collaborative experiences, although the relationship between these transformations and shared cognitive states remains to be evaluated.

\subsection{Future Work}\label{sec:future-work}

Several areas offer promising avenues for future development and exploration. One potential improvement lies in multimodal signal acquisition, where additional biosignals, such as heart rate or galvanic skin response, are incorporated to enhance interactivity. These signals can provide complementary information that could result in more nuanced, personalized adaptations and deeper insights into the user’s emotional and cognitive state. By combining these physiological metrics with EEG data through multimodal fusion techniques, the system can map specific, measurable indicators of affective state to precise adjustments in cinematic content. Such multimodal inputs would enhance the robustness and precision of bio-adaptive systems, allowing for a more holistic interaction that better captures the user's embodied experience.\\

Controlled user studies comparing active and sham EEG conditions are needed to assess user experience, engagement, emotional resonance, and narrative immersion, and to determine whether users can learn to intentionally regulate the EEG features that control visual transformations. Controlled trials using questionnaires, behavioral metrics, and physiological measurements will provide quantitative benchmarks for true bio‑adaptive impact. For example, researchers could examine: (1) viewer engagement via objective measures such as eye tracking; (2) users’ perceptions of control and their beliefs about the narrative being shaped by their mental states through structured questionnaires; and (3) the variability in story interpretations as an indicator of the meaningfulness of EEG-driven augmentations. Insights from these studies will guide iterative refinement of feature extraction, signal processing, and prompt design—advancing both the adaptive cinematic experience and laying the groundwork for future neurofeedback applications targeting cognitive and emotional states.\\

In addition, the exploration of new modalities beyond visual content holds significant potential to further enrich the experience. The integration of audio or tactile feedback could create multi-sensory environments that deepen immersion and engagement. With the rise of new generative models across modalities, our understanding of human-AI relations might gain from the modeling of multimodal embodied interactions, where the user's cognitive, emotional, and sensory states converge to shape the adaptive media experience. \\

Finally, advancements in brain decoding from EEG data present an opportunity to increase the level of user agency in these systems. EEG foundation models offer one possible route for exploring richer signal representations~\citep{cui_neuro_2024}. Future studies could examine whether these representations support more informative mappings to visual transformations and greater perceived control. This would allow for a more accurate and sophisticated adaptation of the media, leading to a stronger sense of connection and co-creation between the user and the AI system.

\subsection{Broader Impact and Ethical Considerations}

By incorporating creative biofeedback, we aim to enhance traditional forms of interaction with cinematic media through adaptivity, rather than replace them. Relying on a human-AI synergistic relationship, these systems may support users' sense of agency during activities such as movie watching. The incorporation of biosignals to augment cinematic content enables a seamless integration of bio-adaptive interfaces and generative visual transformation, creating a dynamically adapted presentation of existing cinematic content that evolves in real time based on the viewer's cognitive state.\\

However, the implementation of such systems raises important ethical considerations around authorship, agency, and data ownership. Clarifying ownership of AI-generated outputs and ensuring compliance with copyright laws for modified media are vital to prevent misuse or legal disputes over creative ownership, and the responsible integration of adaptive, neuro-driven systems into artistic and commercial contexts will require clear ethical frameworks that evolve in line with technological advancements.\\

Moreover, issues around informed consent and data privacy must be thoroughly addressed. Given that the system directly monitors and uses neural activity, users must fully understand how their brain data is being processed, stored, and used. Transparency and security in data handling are paramount, as are mechanisms for users to retain control over their bio-data. Ethical considerations around mental health also play a role, as the real-time reflection of neural states could potentially trigger unintended psychological responses. Safeguards should be in place to protect vulnerable populations and ensure that user well-being remains a priority.

\section{Conclusion}\label{sec13}

The integration of real-time brain-driven interactions with generative AI opens new pathways for embodied co-creation in media. Our system demonstrates the potential of using brain states to modulate visual outputs in real time, contributing to the emerging field of adaptive cinema. By combining both manual and dynamically generated prompts, ReTiNA-Cinema demonstrates how bio-adaptive control can operate not only on fixed stylistic contrasts but also on evolving, context-aware aesthetic spaces, further blurring the boundary between consumption and co-creation.\\

This work lays the groundwork for future explorations into multimodal, bio-adaptive experiences, where content responds to various physiological signals in real time, with the potential to deepen engagement and personalization. As technology advances, the prospect of creating media that is merely consumed but co-created becomes increasingly attainable. This approach has wide-reaching implications, from enriching entertainment experiences to expanding the possibilities of interactive art. By fostering human-AI synergy through real-time brain-responsive systems, this research paves the way for future investigations of agency and engagement across creative domains.


\bibliographystyle{cas-model2-names}

\bibliography{sn-bibliography-revised}
\onecolumn
 
\appendix

\section{Feature extraction pipeline implemented in \textit{goofi}}
 
\begin{appendixfigure}
    \centering
    \includegraphics[width=\textwidth,height=0.72\textheight,keepaspectratio]{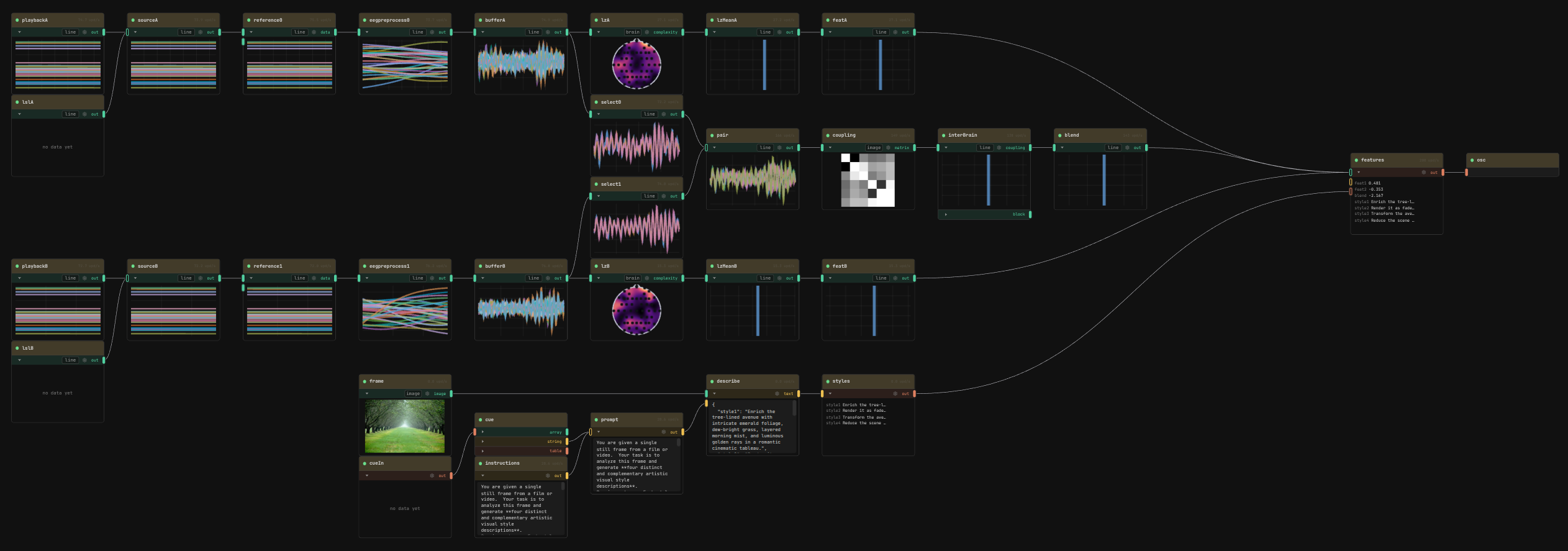}
    \caption{Screenshot of ReTiNA-Cinema's EEG feature extractor implemented in \textit{goofi}. Source: Authors.}
    \label{fig:goofi}
\end{appendixfigure}
 
\section{Generative AI pipeline implemented in \textit{TouchDesigner}}
 
\begin{appendixfigure}
    \centering
    \includegraphics[width=\textwidth,height=0.72\textheight,keepaspectratio]{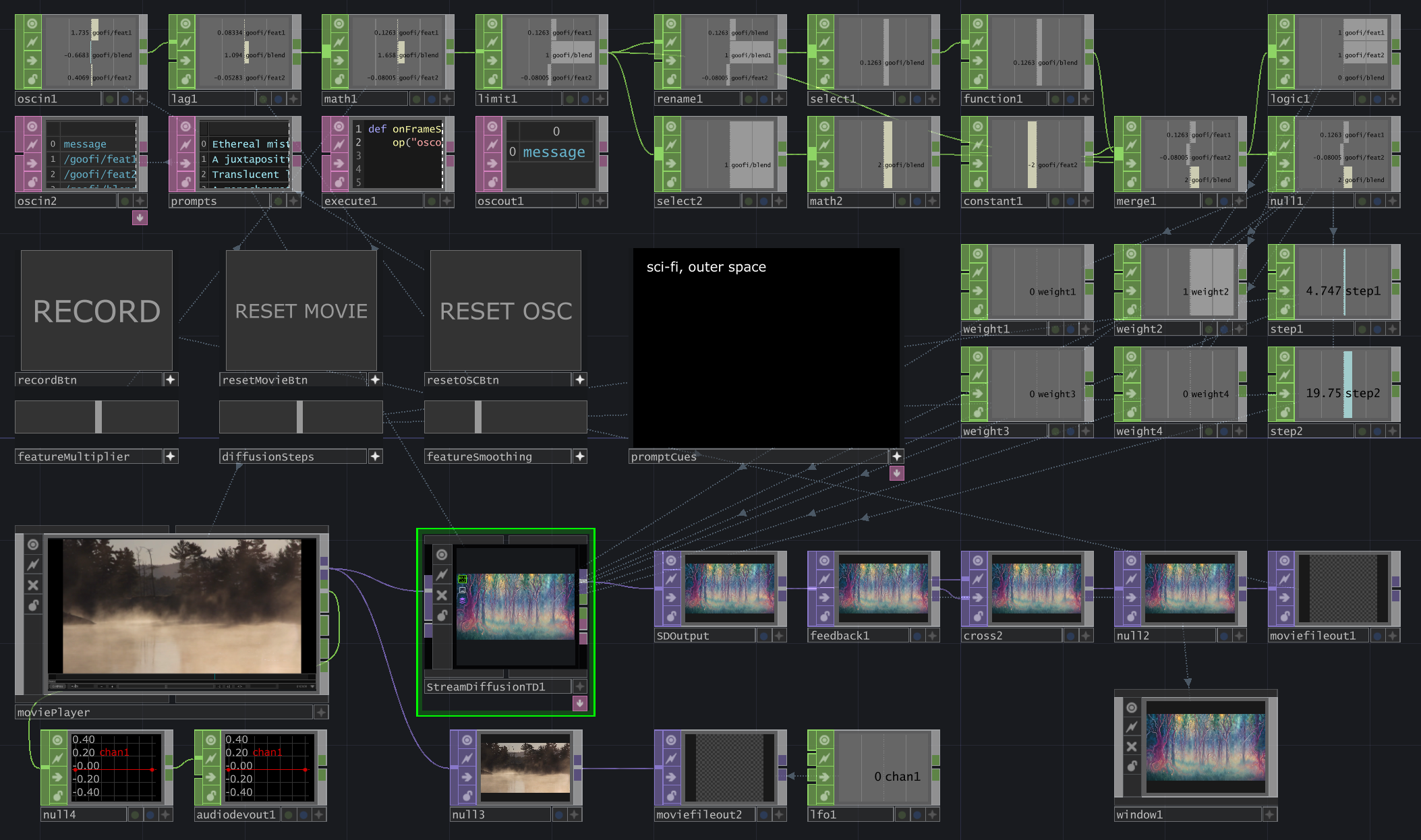}
    \caption{Screenshot of ReTiNA-Cinema's Generative AI Pipeline implemented in \textit{TouchDesigner}. Source: Authors.}
    \label{fig:touchdesigner}
\end{appendixfigure}
 
\clearpage
\section{Context-aware prompt generation}\label{secA2}
 
The following prompt is given to a visual LLM to derive four distinct style prompts from a movie frame:
 
\begin{wrapverb}
You are given a single still frame from a film or video.  
Your task is to analyze this frame and generate **four distinct and complementary artistic visual style descriptions**.  
 
Requirements:  
- Each style must be described in **rich cinematic and artistic language**.  
- Each description must be **maximum 25 words**.  
- The four styles must be **orthogonal/complementary**: they must differ in appearance, mood, emotion, and character.  
- Avoid redundancy; emphasize unique artistic perspectives (e.g., lighting, lens, palette, texture, atmosphere, framing).  
- **Style 1 must always be constructive** (adding styles, details, complexity).  
- **Style 4 must always be deconstructive** (reducing complexity, minimalism).  
- **Styles 2 and 3 are unconstrained**.  
 
Your output must be **only JSON** in the following structure (no extra text, no explanations):  
 
{
  "style1": "description of first style (max 25 words, constructive)",
  "style2": "description of second style (max 25 words, unconstrained)",
  "style3": "description of third style (max 25 words, unconstrained)",
  "style4": "description of fourth style (max 25 words, deconstructive)"
}
 
Always respond strictly in JSON format. Do not include any text outside the JSON object.  
When generating the four style descriptions, incorporate the following styles: <user defined style cues>
\end{wrapverb}
 
 
 
\end{document}